\documentclass[RNAAS]{aastex631}

\usepackage{url}

\begin{document}

\title{Repeating Outbursts from the Young Stellar Object Gaia23bab (=~SPICY~97589)}

\correspondingauthor{Michael A. Kuhn}
\email{m.kuhn@herts.ac.uk}

\author[0000-0002-0631-7514]{Michael A. Kuhn}
\affiliation{Centre for Astrophysics Research, University of Hertfordshire, College Lane, Hatfield, AL10~9AB, UK}

\author[0000-0002-8109-2642]{Robert A. Benjamin}
\affiliation{Department of Physics, University of Wisconsin-Whitewater, Whitewater, WI, 53190, USA }

\author[0000-0002-0406-076X]{Emille E. O. Ishida}
\affiliation{Université Clermont Auvergne, CNRS/IN2P3, LPC, F-63000 Clermont-Ferrand, France}

\author[0000-0001-7207-4584]{Rafael S. de Souza}
\affiliation{Centre for Astrophysics Research, University of Hertfordshire, College Lane, Hatfield, AL10~9AB, UK}

\author[0000-0002-8560-4449]{Julien Peloton}
\affiliation{4IJCLab, Universit\'e Paris-Saclay, CNRS/IN2P3, Orsay, France}

\author[0000-0002-8178-2942]{Michele Delli Veneri}
\affiliation{Instituto Nazionale di Fisica Nucleare Section of Naples, Via Vicinale Cupa Cintia, 26, 80126 Naples, Italy}

\begin{abstract}

The light curve of Gaia23bab (=~SPICY~97589) shows two significant ($\Delta G>2$~mag) brightening events, one in 2017 and an ongoing event starting in 2022. The source's quiescent spectral energy distribution indicates an embedded ($A_V>5$~mag) pre-main-sequence star, with optical accretion emission and mid-infrared disk emission. This characterization is supported by the source's membership in an embedded cluster in the star-forming cloud DOBASHI~1604 at a distance of $900\pm45$~pc. Thus, the brightening events are probable accretion outbursts, likely of EX Lup-type.
\end{abstract}


\section{Introduction} 

Accretion bursts and outbursts from young stellar objects (YSOs) vary in their amplitudes, durations, and spectral properties \citep[e.g.,][]{Herbig1989,Connelley2018,Guo2021,Giannini2022,Fischer2022}. Historically, the scarcity of confirmed events provides challenges for categorization. Nevertheless, bursts and outbursts detected by recent time-domain surveys are beginning to fill out the burst parameter space, and catalogs of YSOs can assist with the quick identification of interesting sources. 

Here we discuss a Gaia alert \citep{Hodgkin2021} for an ongoing outburst from a star in the Spitzer/IRAC Candidate YSO (SPICY) catalog \citep{SPICY}. SPICY was created using data from Spitzer's Galactic midplane surveys, including GLIMPSE \citep{Benjamin2003,Churchwell2009} and related programs, and used a combination of spectral energy distribution (SED) fitting and statistical learning to identify sources with mid-infrared excess indicative of circumstellar disks or envelopes. 

\section{The outburst} 

The Gaia alert, Gaia23bab\footnote{\url{http://gsaweb.ast.cam.ac.uk/alerts/alert/Gaia23bab/}} \citep{2023TNSTR.493....1H}, was issued on 6 March 2023 and noted brightening of SPICY~97589 (19$^h$\,04$^m$\,26.$\!^s$69 +04$^\circ$\,23$^\prime$\,57.$\!^{\prime\prime}$3). Notably, the Gaia $G$-band light curve exhibits two bursts (Figure~1a). The first peaked in mid-2017 with brightening of $\Delta G \approx 2.2$~mag and a 1.1-year duration. During the second, ongoing outburst, the star has brightened from its quiescent magnitude of $G\approx18.7$~mag to 16.3~mag, with most of the brightening happening since April 2022. The ZTF survey \citep{Bellm2019,Graham2019} has detected this brightening in the $r$ band from 19.5~mag to 16.7~mag between July~2022 and Mar~2023 (Figure~1b), and a similar pattern is seen in the ZTF $g$ band. The associated ZTF alert is ZTF22abejrab \citep{Fink}. The source is extremely red (Figure~1c); however, the epoch Gaia BP/RP spectra show the source becoming bluer during both outbursts. The amplitude of the $r$-band rise ($\Delta r \approx 2.8$~mag) meets the 2.5~mag threshold for an outburst \citep{Fischer2022}. 

\begin{figure}[!ht]
    \includegraphics[width=1\textwidth]{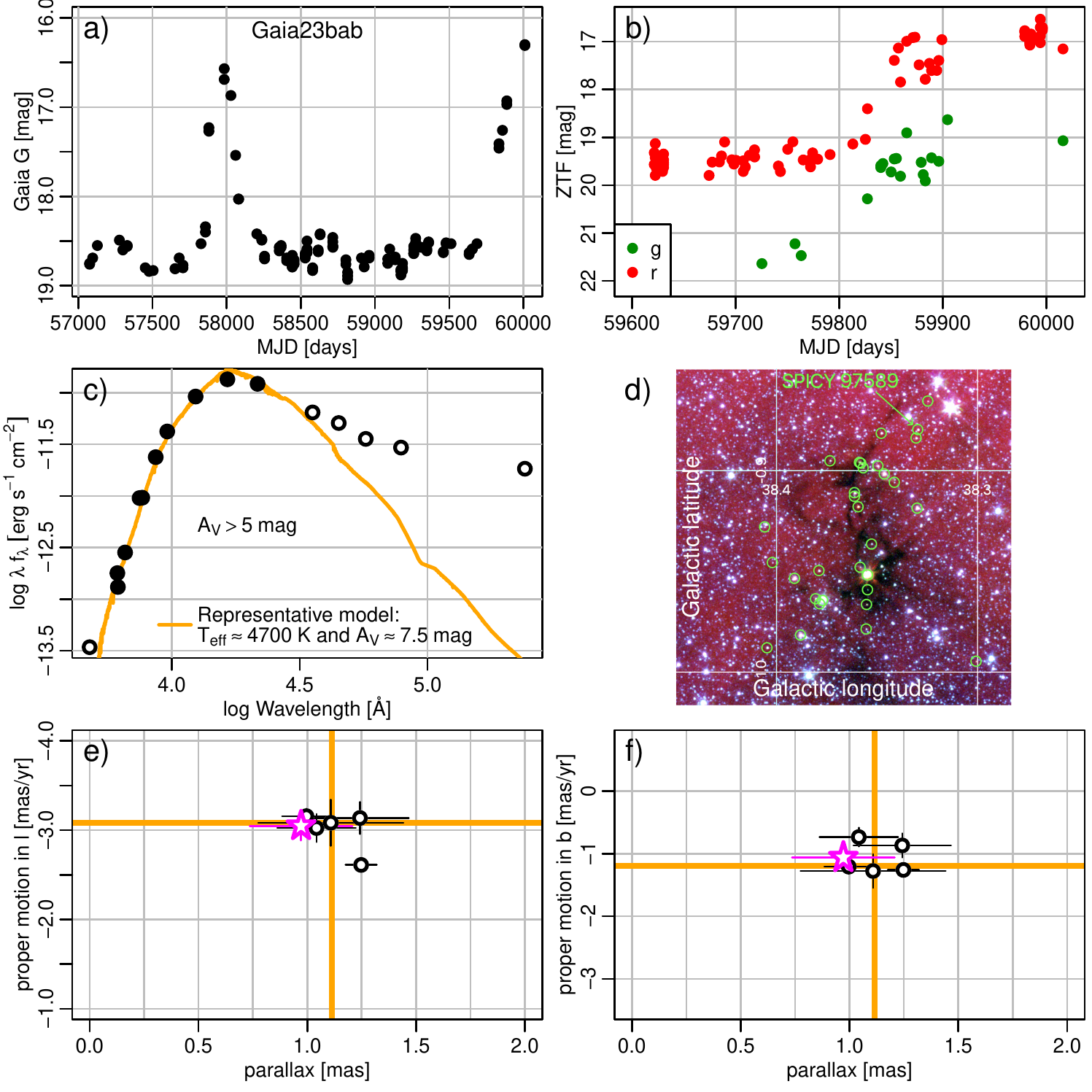}
    \caption{a) Gaia $G$-band light curve for Gaia23bab (= SPICY~97589). b) ZTF $g$ and $r$-band light curves. c) Photometric measurements of the source during quiescence (circles). Filled circles were used for SED fitting, while open circles exhibit excess emission and were excluded from fitting. The orange curve is a representative reddened stellar photosphere model. d)~Spitzer/IRAC 3.6 (blue), 4.5 (green), and 8.0 (red) $\mu$m image of the cloud DOBASHI~1604 and cluster G38.3-0.9. YSOs from SPICY are circled in green. e-f) Parallaxes and proper motions of YSOs in G38.3-0.9 (black circles), with SPICY~97589 indicated by the magenta star. The orange bars indicate the cluster's estimated parallax and mean proper motions. In all panels (a-f), error bars are excluded if they are smaller than the point sizes.}
    \label{fig}
\end{figure}

\clearpage
\section{The star}

The properties of the YSO during quiescence may be inferred from optical and infrared photometric survey data (Figure~1c), including Pan-STARRS \citep{Chambers2016,Flewelling2020}, IPHAS \citep{Drew2005}, 2MASS \citep{Skrutskie2006}, GLIMPSE, and MIPSGAL \citep{Carey2009,Gutermuth2015}. 
To model the star's SED, we used the BOSZ-Kurucz stellar atmosphere model \citep{Bohlin2017} with the \citet{Cardelli1989} extinction law for the optical/near-infrared and \citet{Wang2015} for the mid-infrared. The SED was fit using maximum-likelihood estimation, with an error parameter accounting for stellar variability \citep{Kuhn2023}. Degeneracy between temperature and extinction means that the SED does not strongly constrain the star's effective temperature. Nevertheless, it places a lower limit for the extinction of $A_V>5$~mag. The star's $J-H=1.2$~mag and distance-corrected $M_J=4.3$~mag are consistent with the pre--main-sequence.

For all model fits, excesses are observed in the $g$-band (likely accretion excess) and the mid-infrared bands (likely disk emission). The spectral index of SPICY~97589's mid-infrared SED between the 4.5 and 24~$\mu$m bands is $\alpha=-0.62$, suggesting that the source is a Class~II YSO.

\section{The cluster}

SPICY~97589 is a member of a small cluster of $\sim$30 candidate YSOs (Figure~1d) designated G38.3-0.9 \citep{SPICY}. This group appears partially embedded in a small infrared-dark cloud DOBASHI~1604 \citep{Dobashi2011}, with a mass of $\sim$1800~$M_\odot$ and $v_\mathrm{lsr}=16.9\pm1.4$~km~s$^{-1}$ obtained from CO observations
 \citep{Roman-Duval2010}. The cluster also includes IRAS~19022+0418, which has been classified as an H\,{\sc ii} region by \citet{Urquhart2011}, who also find $V_\mathrm{lsr} = 16.9\pm1.1$~km~s$^{-1}$ from NH$_3$ emission lines.  

Six of the stars in G38.3-0.9, including SPICY~97589, have Gaia DR3 astrometry of sufficient quality to assess membership \citep{Gaia2016,Gaia2021}, i.e., renormalized unit weight errors $RUWE < 1.4$ and parallax uncertainties $<0.4$~mas~yr$^{-1}$. We apply the parallax zero-point corrections from \citet{Lindegren2021}. The six G38.3-0.9 members are significantly more tightly clustered in parallax and proper-motion space than expected for randomly selected stars (Figure~1e,f), indicating they are physically associated. Using the methods from \citet{Spur}, we calculate a cluster parallax of $\varpi=1.114\pm0.056$~mas ($d=900\pm45$~pc) and mean proper motions $\mu_{\ell^\star}=-3.08\pm0.07$~mas~yr$^{-1}$ and $\mu_b = -1.20\pm0.07$~mas~yr$^{-1}$, for Galactic longitude and latitude respectively. 

In a Galactic context, the distance of G38.3-0.9 places it within the ``Split,'' a multi-kpc long molecular filament containing several star-forming complexes, including the Sco-Cen association, the Aquila Rift, and the Serpens Molecular Clouds \citep{Lallement2019}. We calculate Galactocentric velocities of $v_\phi = 237$~km~s$^{-1}$ (azimuthal), $v_r=-0.6$~km~s$^{-1}$ (radial), and $v_z=2.5$~km~s$^{-1}$ (vertical). These imply a nearly circular Galactic orbit, with the \citet{Reid2019} rotation curve yielding an identical azimuthal velocity at the $R=7.47$~kpc Galactocentric radius of this cloud.

\section{Conclusion}

This analysis provides strong evidence for the youth of SPICY~97589, including excess emission in the blue and mid-infrared, resembling an accreting YSO, and astrometry indicating membership in an embedded star-forming region. The year-long duration of the 2017 event and the repeated bursting behavior is typical of EX Lup-type objects (EXors). However, a more complete characterization of these outbursts will require spectroscopic follow-up.   

\begin{acknowledgments}
This work is based on data from ESA’s Gaia mission, processed by DPAC, funded by national institutions, particularly those participating in the Gaia Multilateral Agreement.
\end{acknowledgments}

\end{document}